\documentclass[prl,preprint,showpacs,superscriptaddress,nofootinbib,floatfix]{revtex4}
\usepackage{latexsym}
\usepackage{dcolumn}
\usepackage{epsfig}
\RequirePackage{graphicx}

\newcommand{\ba}{\begin{eqnarray}}

\newcommand{\ea}{\end{eqnarray}}
\newcommand{\be}{\begin{equation}}
\newcommand{\ee}{\end{equation}}

\newcommand{\eq}[1]{Eq.\,(\ref{#1})}

\newcommand{\lessabout}{\raisebox{-.6ex}{\ $\stackrel{<}{\sim }$\ }}
\newcommand{\greaterabout}{\raisebox{-.6ex}{\ $\stackrel{>}{\sim }$\ }}

\newcommand{\pbarp}{\mbox{$\bar p p$\ }}

\newcommand{\WP}{W_{\rm P}}

\def\bea{\begin{eqnarray}} 
\def\eea{\end{eqnarray}} 
\begin{document}

\preprint{ANL-HEP-PR-06-79}
\preprint{Brown-HET-1473}
\preprint{Northwestern-HEP-1251}

\title{Small $x$ Behavior of Parton Distributions from 
the Observed Froissart Energy Dependence of the Deep Inelastic Scattering 
Cross Sections}

\author{M.~M.~Block}
\affiliation{Department of Physics and Astronomy, Northwestern University, 
Evanston, IL 60208}
\author{Edmond~L.~Berger}
\affiliation{High Energy Physics Division,
Argonne National Laboratory,
Argonne, Illinois 60439}
\author{Chung-I Tan}
\affiliation{Physics Department, Brown University, Providence, RI 02912} 
\date{\today}

\begin{abstract}
We fit the reduced cross section for deep-inelastic electron scattering data 
to a three parameter $\ln^2 s$ fit, 
$A +\beta\ln^2(s/s_0)$, 
where $s=\frac{Q^2}{x}(1-x)+m^2$, and $Q^2$ is the virtuality of the 
exchanged photon. Over a wide range in $Q^2$ (   
$0.11\le Q^2\le 1200$ GeV$^2$) all of the fits  satisfy the logarithmic energy dependence of 
the Froissart bound. We can use these results to extrapolate to very 
large energies and hence to very small values of Bjorken $x$ --- well 
beyond the range accessible experimentally. As $Q^2\rightarrow\infty$, the 
structure function $F_2^p(x, Q^2)$ exhibits Bjorken scaling, within 
experimental errors. {We obtain new constraints on the behavior of quark and 
antiquark distribution functions at small~$x$}.
\end{abstract}


\maketitle

{\em Introduction.} Inclusive deep-inelastic lepton scattering (DIS) has 
played a seminal role in particle and nuclear physics, notably for its early 
manifestation of Bjorken scaling~\cite{bjorken}  
and, soon thereafter, the logarithmic scaling violations that are a hallmark 
of perturbative quantum chromodynamics. From the structure functions of DIS, one 
also obtains crucial constraints on the  parton distribution functions 
essential for predictions of short-distance hard-scattering phenomena 
at very high energies, whether at hadron collider facilities or in ultra high 
energy cosmic ray interactions.  The DIS lepton-nucleon cross sections  may also be analyzed 
in complementary fashion as hadronic $Vp$ scattering cross sections, where 
$V = \gamma*$, a virtual photon in the case of electron-nucleon scattering.  
We show explicitly in this Letter that the reduced $\gamma^* p$ total cross sections manifest 
the Froissart $\ln^2 s$ growth~\cite{froissart} with hadronic energy $s$ that also characterizes the 
$\gamma p$, $\pi^{\pm}p$ and $\pbarp$ and $pp$ cross sections at very high energy~\cite{bh}.  
Our fits to 28 different $Q^2$ data sets of $F_2^p(x,Q^2)$ demonstrate that a $\ln^2 s$ 
growth holds for each of the
virtual photon's mass $Q^2$. This observation allows us to obtain a 
new constraint on the Bjorken $x$ dependence of the quark and 
antiquark parton distribution functions at small $x$, viz., they should behave at 
very small $x$ as $\beta\log^2(x_0/x$), for all $Q^2\gg m^2$, where $m$ is the proton 
mass, and $\beta$ and $x_0$ are functions of $Q^2$.  As $x \rightarrow 0$ and $Q^2\rightarrow \infty$, $\beta$ and 
$x_0$ approach constant values $\beta\,',x_0'$.  Only 6 parameters are needed 
to fit 28 data sets, allowing us in principle to extrapolate to very large energies and very low $x$,  
well beyond the experimental range presently accessible, with confidence that we 
understand the $x$ dependence of structure functions and parton distribution functions 
for ``wee'' partons. 

{\em Kinematics.}  In the inclusive process $e p \rightarrow e' X$, the laboratory four-vector 
momentum of the exchanged virtual photon $\gamma^*$ is $ q= (\nu,\vec q)$, with 
negative $q^2\equiv -Q^2$, and the proton laboratory four-vector momentum is 
$ p=(m,0)$.  The invariant $s$, the square of the center of mass (c.m.) 
energy $W$ of the $\gamma^*p$ system, is 
$ 
s\equiv  W^2 = (q+p)^2=2m\nu-Q^2+m^2.\label{Wofnu}
$
The Lorentz invariant variables $x$ and $y$ are defined as  
$ x\equiv \frac{Q^2}{2p\cdot q}=\frac{Q^2}{2m\nu}{\rm { \ \ and\ \ }}
y\equiv \frac{p\cdot q}{p\cdot k},$ 
where $k$ is the incoming electron's four-vector momentum. Thus, 
\be
s=W^2=\frac{Q^2}{x}(1-x) +m^2.\label{W}
\ee

In terms of $x$, $y$ and $Q^2$, the cross section for $e p \rightarrow e' X$ 
may be expressed as 
\ba
\frac{d^2\sigma(x,y,Q^2)}{dxdQ^2}&=&\frac{2\pi\alpha^2}{Q^4x}\left(2-2y+\frac{y^2}{1+R}\right)
F_2^p(x, Q^2) \nonumber\\
&&\equiv\Gamma\times\sigma_{\gamma^*p}^{\rm eff}(x,y,Q^2)\label{sigmaeff}.  
\ea
Here, $F_2^p(x,Q^2)$ is the dimensionless proton structure function,  
$ R\equiv\frac{F_{\rm L}}{F_2-F_{\rm L}},$
with $F_{\rm L}$ being the longitudinal structure function, 
$\alpha$ the fine structure constant, and  
$\Gamma\equiv\alpha(2-2y+y^2)/(2\pi Q^2x)$. 

The quantity measured experimentally, 
$\sigma_{\gamma^* p}^{\rm eff}(x,y,Q^2)$, is a function of 3 variables, $ x, y$ and $Q^2$.
We define the total virtual photon cross section 
$\sigma_{\gamma^* p}^{\rm tot}(x,Q^2)$,  a function of only two 
variables, $x$ and $Q^2$, as 
\ba
\sigma_{\gamma^* p}^{\rm tot}&\equiv&\sigma_{\rm T}(x,Q^2)+\sigma_{\rm L}(x,Q^2)\label{sum}\\
&=&\frac{4\pi^2\alpha}{Q^2}\times \left(\frac{1+4m^2x^2/Q^2}{1-x}\right)F_2^p(x, Q^2)\label{longform}\\
&\approx &\frac{4\pi^2\alpha}{Q^2}\times{F_2^p(x, Q^2)\over 1-x},\quad   Q^2\gg 4m^2x^2.  
\label{simple}
\ea
In \eq{sum}, $\sigma_{\rm T}(x,Q^2)$ and $\sigma_{\rm L}(x,Q^2)$ denote the transverse and 
longitudinal virtual photon cross sections. 

{\em Saturation of the Froissart bound.} High energy cross sections for hadron-hadron scattering 
must be bounded   by $\sigma \sim \ln^2s$, where $s$ is the square of the c.m. energy.   This 
fundamental result is derived from unitarity and analyticity by Froissart~\cite{froissart}.
Saturation of the Froissart bound refers to an energy dependence of the total cross 
section rising no more rapidly than $\ln^2s$. 

It has been shown that the Froissart bound is {\em saturated} at high energies~\cite{bh} in 
$\gamma p$, $\pi^{\pm}p$ and $\pbarp$ and $pp$ scattering, and, as we now will show, 
also in $\gamma^*p$ scattering.

We choose to work in terms of a dimensionless ``reduced'' $\gamma^*p$ cross section,
$\sigma_{\gamma^*p}^{\rm tot}(W, Q^2)/\kappa$, where  
$\kappa\equiv 4\pi^2\alpha/Q^2$.  Using the parameterization of Block and Cahn~\cite{bc}, 
we write the  reduced cross section{\footnote {The parameterization of \eq{sigofW} is the same 
as that used by Block and Halzen~\cite{bh} for their successful fits of hadronic cross 
sections for $\gamma p$, $\pi^{\pm}p$ and $\pbarp$ and $pp$ scattering, except for a transformation 
of variables.}} as 
\ba
\sigma_{\gamma^*p}^{\rm tot}(W, Q^2)/\kappa &=& A+\beta \ln^2{s\over s_0} + cs^{-0.5}.\label{sigofW}
\ea
The 4 coefficients $A$, $\beta$, $s_0$, and $c$ are functions of $Q^2$.
We present fits to 29 data sets published by the ZEUS~\cite{zeus}
collaboration\footnote{In order to avoid possible normalization differences, we have not included 
H1 data~\cite{H1} in our analysis, although preliminary examination of the combined data sets leads 
us to identical physics conclusions.}%
, for $Q^2$ = 0.11, 0.20, 0.25, 0.65, 2.7, 3.5, 4.5, 6.5, 8.5, 10, 12, 15, 18, 22, 27, 
35, 45, 60, 70, 90, 120, 150, 200, 250, 350, 450, 650, 800, and 1200~GeV$^2$.   
Prior to displaying and discussing our fits, we make three observations about the data:  
\begin{itemize}
\item An examination of plots of $F_2^p(x, Q^2)$ versus $x$ for different values of $Q^2$ 
shows that all data sets are compatible with going through a {\em common} point at 
$x\approx 0.09$ and $F_2\approx 0.41$.  We call this common intersection 
the ``scaling''  point $x_{\rm P}$ and examine its significance below.  
\item Inspection of $F_2^p(x, Q^2)$ versus $x$ shows that for $x \le x_P$ 
and $Q^2 \greaterabout 350$ GeV$^2$, the data overlap each other, i.e., within experimental 
uncertainties, they scale  in Bjorken $x$ (see footnote \ref{foot:bjscaling}).
\item When plotted as $\sigma_{\gamma^*p}^{\rm tot}/\kappa$ versus $W$,  
with a logarithmic $W$ axis, the data at all $Q^2$ show similar parabolic shapes for $x\le x_{\rm P}$, 
rising with $W$, with the 
shape parameters depending slowly on  $Q^2$,  with a negligible inverse-power term proportional to $c$.  
\end{itemize}
Introducing $x$ from \eq{W}, defining the parameter $x_0(Q^2)\equiv Q^2/s_0$, 
and setting $c = 0$, we find that \eq{sigofW} becomes, for $x\le x_P$,
\ba
\sigma_{\gamma^*p}^{\rm tot}(x, Q^2)/\kappa \!\!\!&=&\!\!\!\!A+\beta\ln^2\left[ x_0\frac{1-x}{x}+{x_0m^2\over Q^2}\right].
\label{sigofx}
\ea 
For $Q^2\gg m^2$ and $x\le x_P$, we obtain the closed forms 
\ba
\!\!\!\!\!\!\sigma_{\gamma^*p}^{\rm tot}(x, Q^2)/\kappa &=&A+\beta\ln^2\left[ x_0\frac{1-x}{x}\right],\label{approxsigofx}\\ 
F_2^p(x,Q^2)&=&(1-x)\left(A+\beta\ln^2\left[ x_0\frac{1-x}{x}\right]\right),\label{F2ofx}
\ea
where $A$, $\beta$, and $x_0$ are functions of $Q^2$.

In order to bolster the above observations quantitatively, we made $\chi^2$ fits of \eq{sigofx} to 29 sets of the Zeus data, i.e., 
using the 
 $\ln^2s$ parameterization of  \eq{sigofW} with $c=0$, with the added requirement that each 
curve go through its appropriate ($W_{\rm P}(Q^2),\sigma_{\gamma^*p}^{\rm tot}(W_{\rm P}, Q^2)/\kappa$) 
scaling point, where 
\be 
  W_{\rm P}(Q^2)\equiv W(Q^2,x_{\rm P})
=\sqrt{\frac{Q^2}{x_{\rm P}}(1-x_{\rm P})+m^2}.
\ee 
We fit only those data with $x\le x_{\rm P}$, corresponding to $W > W_P$, 
which is our definition of high energy data.  We treat the 5 highest $Q^2$ data sets, 
$Q^2=350,450,650,\\ 800$ and 1200 GeV$^2$, as one set, since their $F_2^p(x)$ data points 
overlap. From fits to individual sets of points with common 
$Q^2$, we find that 
over the range\footnote{For $Q^2> 1200$ GeV$^2$, there are essentially no ZEUS data with 
$W > W_{\rm P}(1200$ GeV$^2$), so we stop our fits at $Q^2=1200$ GeV$^2$.} 
$0.11\le Q^2\le 1200$ GeV$^2$,  and for energies $W\ge W_{\rm P}$, each data set with 
a common $Q^2$ is fit 
satisfactorily with the  3 parameters $A, \beta$ and $s_0$, with $A$  
constrained by 
\ba
A&=&\sigma_{\gamma^*p}^{\rm tot}(W_{\rm P}, Q^2)/\kappa-\beta\ln^2 {W_{\rm P}^2\over s_0}. \label{A}
\ea

\begin{figure}
\begin{center}
\mbox{\epsfig{file=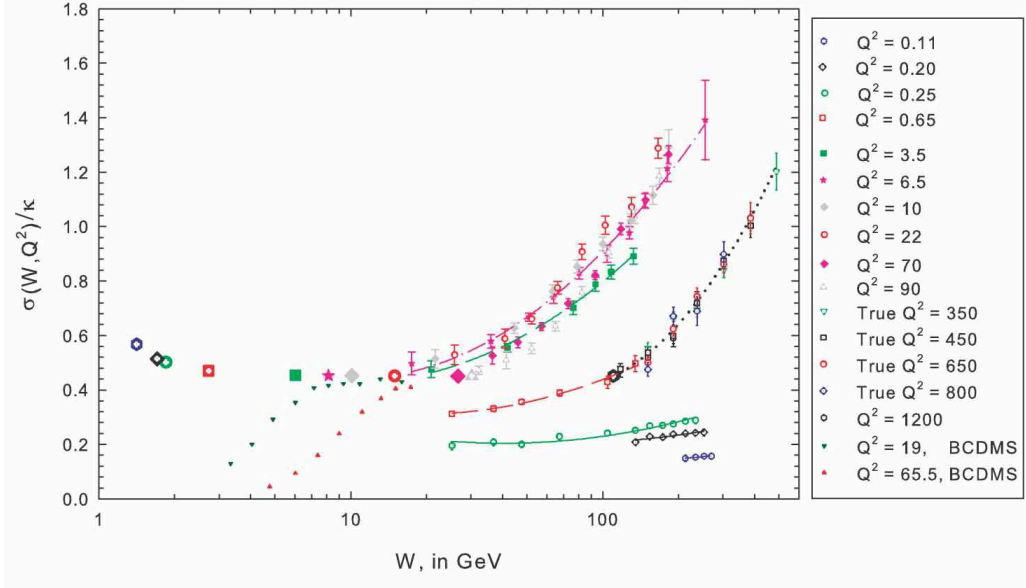,width=5.4in%
,bbllx=0pt,bblly=208pt,bburx=533pt,bbury=513pt,clip=%
}}
\end{center}
\caption[]{
The reduced cross section $\sigma_{\gamma^*p}^{\rm tot}(W, Q^2)/\kappa$  vs. $W$, 
in GeV, with $\kappa=4\pi^2\alpha/Q^2$. The high energy Zeus~\cite{zeus} data plotted 
here  have $x\le x_{\rm P}$. The curves are $\ln^2s$ fits,
$\sigma_{\gamma^*p}^{\rm tot}/\kappa=A + \beta\ln^2 (W^2/s_0)$,  
forced to go through the scaling point $x_{\rm P}, F_2^p(x_{\rm P})$. 
The 11 {\em large} symbols, $\WP(Q^2)$, correspond to the  common intersection (scaling) 
point shown in  Fig. \ref{fig:F2ofxfit}. The very tiny points, data from the BCDMS 
collaboration~\cite{bcdms}, are not used in the fits. The data labeled ``true $Q^2=$'' are fit 
as if they had been part of the   $Q^2=1200$ GeV$^2$ set.
}
\label{fig:sigmaoverkappafit}
\end{figure}

Our results are illustrated in Fig.~\ref{fig:sigmaoverkappafit}, where we plot the 
reduced cross section vs. the c.m. energy $W$ for the Zeus~\cite{zeus} high 
energy data, together with our constrained fits. We use their published values 
of $F_2^p(x)$ and for errors, their statistical and systematic errors taken in 
quadrature.  
All of the data%
\footnote{Although not shown in Fig. \ref{fig:sigmaoverkappafit} to avoid loss of clarity, 
we have also made constrained fits for $Q^2$= 2.7, 4.5, 8.5, 12, 15, 18, 27, 35, 45, 60, 
120, 150, 200 and 250 GeV$^2$, with similar results, a total of 25 fits to independent data sets 
at different $Q^2$. We found a total $\chi^2=113.56$ for 164 degrees of freedom, 
corresponding to $\chi^2/d.f.=0.692$.  For the 5 combined data sets 
at the highest $Q^2$, we found $\chi^2=26.99$ for 27 degrees of freedom, {\em a posteriori} 
justifying our combining them into a single set labeled 1200.\label{foot:bjscaling}}%
\  agree well 
with a $\ln^2s$ parameterization over this wide  $Q^2$ range. Also shown in Fig.~\ref{fig:sigmaoverkappafit} as large bold points 
are the 11 different values of $W(x_{\rm P})$ that correspond to the scaling point 
 $x_{\rm P}=0.09,\ F_2^p(x_{\rm P})=0.41$.     
 
\begin{figure}
\begin{center}
\mbox{\epsfig{file=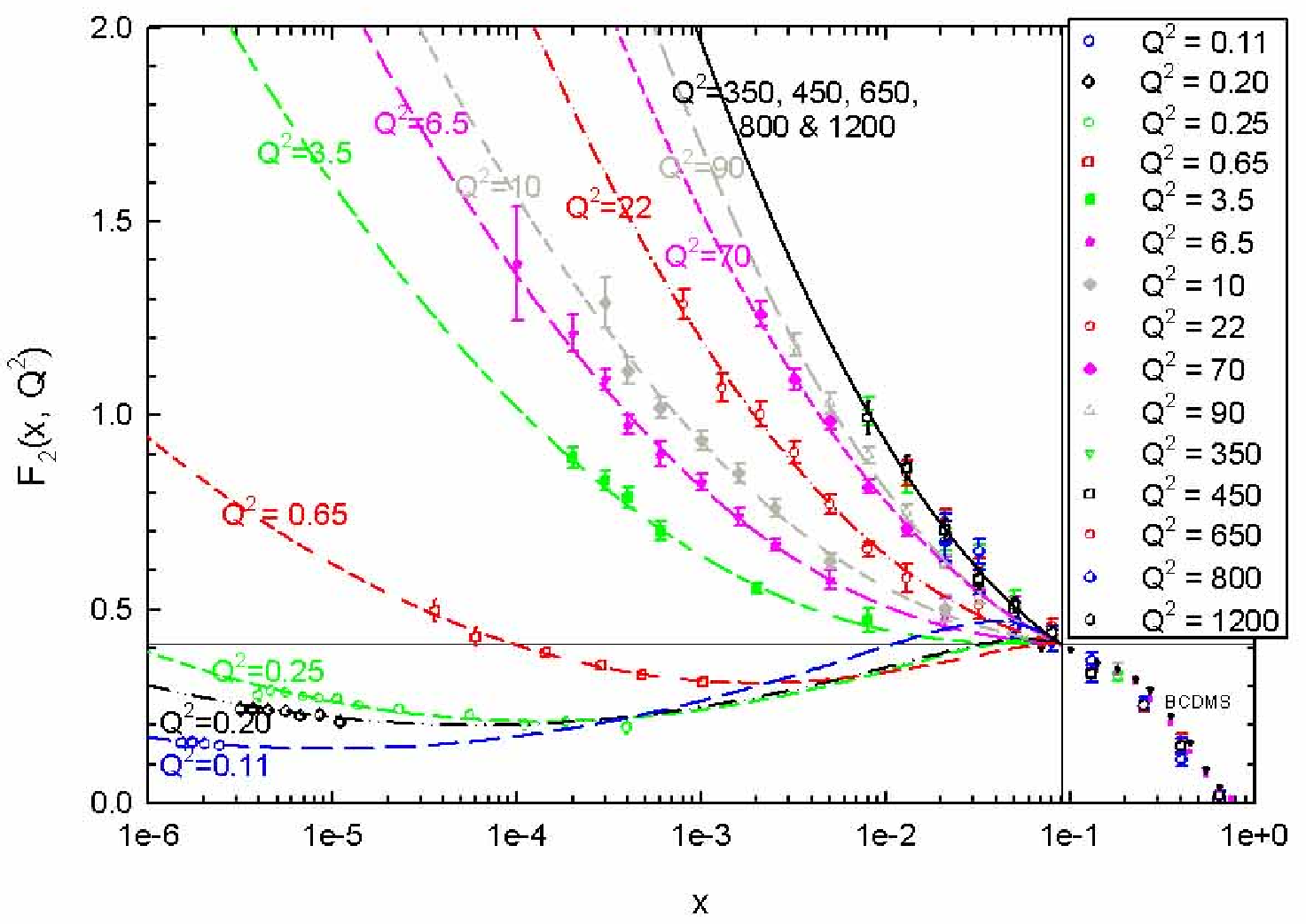,width=5.4in%
,bbllx=0pt,bblly=290pt,bburx=425pt,bbury=600pt,clip=%
}}
\end{center}
\caption[]{ 
Fits to the proton structure function data, $F_2^p(x, Q^2)$ vs. $x$, for 15 values of $Q^2$.  
The data are those of Fig.~\ref{fig:sigmaoverkappafit}. The curves are the $\ln^2s$ fits of 
Fig.~\ref{fig:sigmaoverkappafit}, converted to the $F_2^p-x$ plane. The tiny points 
are  BCDMS~\cite{bcdms} data. The vertical and horizontal straight lines intersect at the 
scaling point $x_{\rm P}=0.09,F_2^p(x_{\rm P})=0.41$. The data  for $Q^2=350,450,650,800$ and 
1200 GeV$^2$ are all fit together with a single curve. \label{fig:F2ofxfit}}
\end{figure}

The agreement with experiment, over the broad range of $Q^2$ investigated here, 
is seen perhaps more clearly in the plot of $F_2^p(x)$ vs. $x$, shown in Fig.~\ref{fig:F2ofxfit}. 
This figure shows the same  $Q^2$ sets as in Fig.~\ref{fig:sigmaoverkappafit}, 
along with the same 11 constrained $\ln^2s$ fits of the form specified in \eq{sigofx}, 
with $A$ constrained by \eq{A}.  Careful examination of Fig.~\ref{fig:F2ofxfit} shows that the data are in 
excellent agreement with the hypothesis that all of the widely different $Q^2$ fits intersect at the  point $x_{\rm P}=0.09,F_2^p(x_{\rm P})=0.41$, i.e., it's  
is a very good scaling point, being satisfied by all of the ZEUS~\cite{zeus} 
sets as well as the BCDMS collaboration points~\cite{bcdms}.   


Having a scaling point gives us a universal anchor point ---an {\em analyticity constraint}~\cite{block}---
for the fits to the different sets of values of $Q^2$.  As shown in \eq{A}, only 2 parameters, 
$\beta$ and $x_0\equiv Q^2/s_0$,  
are now required for the fit to each $Q^2$ data set. This  reduction in the number of parameters 
constrains their values, making the fit uncertainties considerably smaller, e.g., the 
fractional error in $F_2^p$  due to parameter uncertainties  is $\sim 5$\% , for both 
$x=10^{-6}$, $Q^2=3.5 {\rm \ GeV}^2$ and $x=10^{-3}$, $Q^2$= {\rm \ 1200} GeV$^2$.

Although not shown explicitly here, we find that the coefficients $\beta$ and $x_0$ 
are smoothly varying  functions of $Q^2$, both being parameterized adequately 
as 
\ba
\ln x_0&=&\ln x_0'+{{a_1}\over(Q^2)^{a_2}},\label{x_0} \\
\beta&=&\beta\,'-{{b_1}\over(Q^2)^{b_2}},   
\label{beta}
\ea
where the constants $\beta\,'$ and $x_0'$ are given by
\be 
\beta\,',x_0'=\lim_{Q^2\rightarrow\infty}\beta(Q^2),x_0(Q^2).
\ee 
The 6 constants $\beta\,',x_0',a_1,a_2,b_1 $ and $b_2$ are 
parameters that can be obtained in a global fit. In a truly global fit, in addition 
to these 6 constants we need the two scaling point coordinates $x_{\rm P},F_2^p({x_{\rm P})}$, i.e., a total 
of 8 parameters, in order to fit all the available data on $F_2^p(x, Q^2)$. This 
Letter demonstrates the feasibility of such a fit.

Attention may be drawn to the fits for $Q^2$ = 0.11, 0.20, 0.25 and 0.65~GeV$^2$ in 
Figs.~\ref{fig:sigmaoverkappafit}~and~\ref{fig:F2ofxfit}.  These curves 
do not show different shapes in $W-\sigma/\kappa$ space from any of their counterparts 
at larger $Q^2$---since they also are parabolic in shape---but they exhibit more 
structure than the other curves when plotted  in $F_2^p-x$ space.  This shape difference 
is explained by their very low values of $Q^2$. The term $1+4m^2x^2/Q^2$ in \eq{longform} 
becomes substantially greater than unity as $x \rightarrow x_{\rm P}$, and in the 
relation $\frac{Q^2}{x}(1-x) =W^2-m^2$ of \eq{W}, 
$\frac{Q^2}{x}(1-x)\approx 2m(W-m)$ for small  $Q^2$ and 
$x \lessabout  x_{\rm P}$; these terms strongly influence the transformation from 
$\sigma^{\rm tot}_{\gamma^{*p}}/\kappa-W$ space to $F_2^p-x$ space when $Q^2$ is very small.  
We find that one closed form  fits {\em all} of the $Q^2$ data for $x\le x_{\rm P}$, i.e., 
\ba
F_2^p(x,Q^2)&=&\frac{1-x}{1+4m^2x^2/Q^2}\times\nonumber\\
&& \!\!\!\!\!\!\!\!\!\!\!\!\!\!\!\!\!\!\!\!\!\!\!\!\left\{A+\beta\ln^2\left[ x_0\frac{1-x}{x}\left( 1+\frac{m^2}{Q^2}\frac{x}{1-x}\right)\right]\right\}.  
\ea
\\
{\em Implications for parton distribution functions.}  The structure function $F_2^p(x,Q^2)$ 
has a simple interpretation in the parton model in which the scattering from 
the proton is due entirely to the scattering from its individual constituents.  Only the 
quarks and antiquarks couple directly to the electroweak 
current carried by the virtual photon $\gamma^*$ in DIS.  The DIS cross section may then 
be expressed in terms of the probabilities $q_{f/h}(x)$ of finding a quark, and 
${\bar{q}}_{f/h}(x)$ of finding an antiquark, of flavor $f$ and fractional momentum $x$ 
in hadron $h$, times the cross section for the elastic scattering of that parton. 
As a consequence of gluon radiation in quantum chromodynamics (QCD), the quark, antiquark, 
and gluon densities become functions of $Q^2$ as well as $x$, e.g., $q_{f/h}(x) \rightarrow 
q_{f/h}(x, Q^2)$.  

The precise expression for $F_2(x, Q^2)$ in terms of parton distributions depends on the 
choice of the factorization scheme in QCD.  The most commonly used schemes are the 
DIS and the modified minimal subtraction ${\overline{ \rm MS}}$ schemes. In the DIS scheme, 
all higher-order contributions to the structure 
functions $F_2(x,Q^2)$ are absorbed into the distributions of the quarks and 
antiquarks~\cite{altarelli1979}, and we may write, to all orders in QCD,  
\be
 F_2(x,Q^2) =\sum_f e_f^2 x (q^{\rm DIS}_{f/h}(x,Q^2) + {\bar q}^{\rm DIS}_{f/h}(x,Q^2)),  
\label{qcd}
\ee
where $e_f$ is the fractional electric charge of flavor $f$.
The superscripts ${\rm DIS}$ indicate that the distributions are those defined 
in the DIS scheme. In QCD, gluon radiation removes momentum at large $x$, decreasing the 
value of $F_2^p(x, Q^2)$ as $Q^2$ grows, and builds up the 
quark and antiquark distributions at small $x$, leading to the qualitative expectation 
of the {\em scaling} point that we observe in $x$ at which $F_2^p(x_P, Q^2)$ is independent of 
$Q^2$.   

The  Froissart $\ln^2(s/s_0)$ energy dependence of the data allows us to 
conclude, for $x\le x_P$ and $Q^2\gg m^2$, that  
\ba
\!\!\!\!\!\!\!\!\! xq^{\rm DIS}_{f/h}(x,Q^2)&\sim &(1-x)\times
\left(A+
\beta\ln^2\left[x_0 \frac{1-x}{x}\right]\right) .
\label{smallx}
\ea 
Moreover, within experimental errors, we find that 
\ba 
xq^{\rm DIS}_{f/h}(x,Q^2\greaterabout 350 {\rm \ Gev}^2)&\approx&xq^{\rm DIS}_{f/h}(x)
\nonumber\\
&&\!\!\!\!\!\!\!\!\!\!\!\!\!\!\!\!\!\!\!\!\!\!\!\!\!\!\!\!\!\!\!\!\!\!\!\!\!\!\!\!\!\!\!\!\!\!\!\!\!\!\!\!\sim(1-x)\times\left(A'+
\beta\,'\ln^2\left[x_0' \frac{1-x}{x}\right]\right) , 
\ea
with $Q^2$-independent values $A'=0.40,\beta\,'=0.050\pm0.008$, and $x_0'=0.28\pm0.01$.

Parton distribution functions of quarks, antiquarks, and gluons are generally extracted 
from a global fit~\cite{cteqmrst} to a wide class of hard scattering data including deep-inelastic 
scattering cross sections.  Typically one begins with assumed parameterizations of 
the $x$ dependence of the distributions at a fixed low value $Q_0$ and then uses 
the perturbative evolution equations~\cite{dglap} of QCD to obtain the $x$ and $Q^2$ 
dependences of these distributions for all $Q > Q_0$.  The non-perturbative form that 
is assumed in the CTEQ parameterization, for example, for these distributions has power 
behavior at small x, 
$
x f(x,Q_0) \sim x^A . 
$
We contrast this power behavior  with the logarithmic 
behavior that is shown by the data,  suggesting  that it is productive 
to redo a global fit program using  logarithmic 
behavior at small $x$.  Over a small range of $x$ one cannot distinguish a 
logarithmic and a power expression with small fractional power 
(e.g., $\sim 0.25$), but the behavior of these expressions is clearly 
different when extrapolated over a very great range.     
    
In conclusion, we demonstrated that we can make simultaneous $\ln^2s$ fits
 saturating the Froissart bound, to 218 ZEUS datum points, with all fits going 
through the same scaling point, strongly constraining  the behavior of $F_2^p(x,Q^2)$ at tiny 
$x$.  These fits are made with only 2 parameters per set of data with common $Q^2$, since 
the analyticity constraint~\cite{block} of \eq{A} on the cross section at the scaling point requires 
only the 2 coefficients, $\beta(Q^2)$, $x_0(Q^2)$.  This additional constraint insures a well 
determined set of fit parameters, and thus an accurate forecast of the very small $x$ behavior.  
Saturation of the Froissart bound allows us to project to very large $W$ and hence to very small 
$x$ with a high degree of confidence in our functional form.  Our demonstration of Froissart 
energy dependence at small $x$ for each $Q^2$ should shed light on various theoretical efforts in 
examining small $x$ physics in QCD and on parton saturation at high density~\cite{smallx}. Our 
program for the future includes a global constrained $\ln^2 s$ fit, 
as described above, to all available $F_2^p$ data, in order to obtain  the necessary parameters 
and their uncertainties needed  to extract accurate parton distribution functions at small $x$. 

\begin{acknowledgments}
 E.L.B. is supported by the U.~S.\ Department of Energy, Division of High 
Energy Physics, under Contract No.\ W-31-109-ENG-38.
C-I.T. is supported in part by the U.~S.~Department of Energy under 
Contract DE-FG02-91ER40688, TASK A.  
E.L.B. and M.M.B. thank the Aspen Center for Physics for its 
hospitality during the writing of this paper. E.L.B. thanks Professor 
Jianwei Qiu for valuable comments, 
and M.M.B.  thanks Professors L. Durand III, S. Gasiorowicz, and 
F. Halzen for providing new insights and most 
valuable discussions.
\end{acknowledgments}

\end{document}